\documentstyle[multicol,aps,epsfig]{revtex} % for multicolumn
\tighten       %Gives single-space (RevTex 3.0)

\begin{document}
%prints PACS numbers in \pacs{ }
\draft

\title{Temperature suppression of STM-induced desorption of hydrogen
on Si(100) surfaces}
\author{C. Thirstrup, M. Sakurai, T. Nakayama}
\address{Surface and Interface Laboratory, RIKEN, Saitama 351,
Japan.}  
\author{K. Stokbro}
\address{Mikroelektronik Centret, Danmarks Tekniske Universitet, 
Bygning 345\o , DK-2800 Lyngby, Denmark.}
\date{\today}
\maketitle
\begin{abstract}
The temperature dependence of hydrogen (H) desorption from 
Si(100) H-terminated surfaces by a scanning tunneling
microscope (STM) is reported for negative sample bias. It is found that
the STM induced H desorption rate ($R$) decreases several orders
of magnitude when the substrate temperature is increased from
300 K to 610 K. This is most noticeable at a bias voltage of
$-7$ V where $R$ decreases by a factor of ~200 for a temperature
change of 80 K, whilst  it only  decreases  by a factor of ~3  at $-5$ V 
upon the same temperature change. The experimental data can be
explained by desorption due to  vibrational
heating by  inelastic scattering via a hole resonance. This
theory predicts a weak  suppression of   desorption  with
increasing temperature due to a decreasing vibrational lifetime, and a
strong bias dependent suppression due to a temperature dependent lifetime of
the hole resonance.
\end{abstract}
\begin{multicols}{2}

\narrowtext

Desorption of hydrogen (H) from a  H-terminated silicon (Si)
surface has recently been studied
actively\cite{BeHiChBe90,LyAv90,LyAbShWaTu94,ShWaAbTuLyAvWa95,ScRaEnHaKo96,AvWaRoShAbTuLy96,HuYa96,ShAv97,FoKaLyAv98,StThSaQuHuMuGr98}, because of
its potential applications to construct novel optical and electronic
devices. For positive sample bias, Shen et al.\cite{ShWaAbTuLyAvWa95} have
shown that below 4 V the STM induced H desorption is related to
vibrational heating\cite{GaPeLu92,WaNeAv93} by inelastic scattering of
tunneling electrons with the Si-H 6$\sigma^*$  resonance, and above 4 V it is
related to direct electronic excitation of the Si-H bond. At
negative sample bias, the desorption mechanism has recently
been explained by the authors\cite{StThSaQuHuMuGr98} to be due to inelastic
scattering of tunneling holes with the Si-H 5$\sigma$ hole resonance
and quantitative agreement between experimental desorption data
and first principles calculations was obtained.

The present paper focuses on the temperature($T$) dependence of the
STM induced H desorption rate($R$) from Si(100)-H(2$\times1$) surfaces in  ultra high vacuum (UHV)  at
negative sample biases from $-10$ V to $-4$ V and temperatures from 300 K
to 610 K. For a given sample bias($V_{\rm b}$) and tunnel current($I$) we measure a
decreasing $R$ with increasing $T$. The effect is
most pronounced at $V_{\rm b}$ =  $-7$ V, 
where $R$ decreases by a factor ~200 for an increase in $T$ of 80 K,
whilst  it  only decreases by a factor ~3 at $V_{\rm b}$ =  $-5$ V upon
the same temperature change. Previous studies of the $T$ dependence of $R$  have been carried out for  positive sample bias and at $T=11$ K, where
$R$  was reported to increase by a factor of
$\sim 300$ compared to room temperature\cite{FoKaLyAv98}. The lower
$R$ with increasing $T$ was explained by  a decreasing
vibrational life time of the Si-H stretch mode with increasing $T$\cite{TuChRaBoLu85,PeAv97}.

In the present study  the $T$ dependence of $R$ has been modelled using a first principles model\cite{StHuThXi98} of  
inelastic scattering of tunneling holes  with
the Si-H 5$\sigma$ hole resonance. Two temperature effects, which are based on the  $T$ dependence of the
lifetime of  vibrational excitations and electronic excitations, respectively, are included in the model. At $V_{\rm b}=-5$ V,  the 
$T$ dependence of $R$ is related to the lifetime of the vibrational
excitation, which is similar to the result for positive $V_{\rm b}$
reported previously by Foley {\it et. al.}\cite{FoKaLyAv98}.  At
$V_{\rm b}=-7$ V the $T$ dependence from this effect is too weak to
explain the experimental data. The strong $T$ dependence of $R$ in
this case is mainly caused by the $T$ dependence of  the
lifetime of the 5$\sigma$ hole resonance, which  gives rise to an additional
strongly bias dependent temperature effect. The combination of a $T$
dependent vibrational lifetime and a $T$ dependent electronic lifetime
accounts very well for the experimental data in the whole bias range
from $-10$ V to $-4$ V. 

The experiments were performed at a base
pressure of $\sim 7 \times 10^{-9}$ Pa using  $n$-type  Si(100) ($N_D=1\times
10^{18}\; {\rm cm}^{-3}$) samples and  a JEOL JSTM-4000XV microscope
with electrolytically sharpened tungsten (W) tips. Samples were 
heated using a direct current flowing between two contacts at each end of 
the samples, the STM tip being positioned in the middle of the two 
contacts. A change in $V_{\rm b}$ caused by the 
current flow was corrected for in the STM bias circuit, and the absolute 
value of $V_{\rm b}$ was verified by measuring the zero point in
current-voltage  curves. The Si(100)-H(2$\times1$)
 surfaces were prepared by standard
procedures\cite{LyAbShWaTu94}. The surfaces are constructed by an array
of parallel  'dimer rows', where the dimer consists of two
dimerized Si atoms each bonded to a single H atom. If one of
the H atoms is desorbed, a Si dangling bond appears and is
observed as a bright spot in filled-state STM images. The
desorption rate ($R$) of  H atoms  is determined by the
following STM parameters: the scanning speed ($s$) of the STM
tip, $V_{\rm b}$, $I$ and $T$. The experiments were carried out at
negative $V_{\rm b}$ between $-10$ V and $-4$ V and at $T$ between 300 K and
610 K. Increasing $T$ above 610 K caused an excessively large
thermal desorption of H atoms and the study is therefore
limited to $T< 610$ K.

 In order to establish the relationship between $R$,  $I$, $V_{\rm b}$ and
$T$, the surface was scanned by the STM tip at various
conditions, and $R$ could be determined from $s$  and the number
of desorbed H atoms identified from the STM images.  Two
different kinds of experimental methods were employed. In one
method  the STM tip raster scanned
areas of 20 nm $\times$ 20 nm on the Si(100)-H(2$\times1$) surface using a fixed value of $V_{\rm b}$, while  $I$
and $s$ were adjusted to yield a suitable countable number of
desorbed H-atoms (typically 100-200 atoms). This method was
used to determine R as function of $I$ for constant $V_{\rm b}$. In
another method, line scans with a fixed $s = 2$ nm/s were
performed at various $V_{\rm b}$, and $I$ was adjusted to yield
desorption of 50\% of the H atoms along a line scan on the
surface. This method was used to determine the relationship
between $V_{\rm b}$ and $I$ for constant $R$.

 Figure~1 shows a filled state STM image recorded at $T=530$ K
after desorption of H from the Si(100)-H(2$\times1$) surface. The bright
line running from the top right to the bottom left is a chain
of Si dangling bonds created by scanning the tip along a line
with $s = 2$ nm/s,  $V_{\rm b} =  -7.0$ V and $I = 8.0$ nA . The width of
the line is approximately equal to the width of one dimer row
and the resolution is similar to that obtained for lithography
at room temperature\cite{StThSaQuHuMuGr98}.

Figure~2 shows $R$  plotted as function of  $I$ for (a) $V_{\rm b} = -7$ V
and (b) $V_{\rm b} = -5$ V and for temperatures $T= 300$ K (squares), 380 K
(crosses) and 450 K (triangles). We have made least  squares fits to the data of a power-law
dependence of $R$ upon $I$, $R=R_0 (I/I_{\rm des}) ^{\alpha}$, where
$I_{\rm des}$ is the tunnel current  corresponding to $R_0= 4$
s$^{-1}$.  At $V_{\rm b} =  -7$ V  the least-squares fitted 
 parameters are: $\alpha=5.7\pm0.7$,
$4.7\pm0.5$ and $4.6\pm0.7$, and $I_{\rm des} =
0.85$ nA, $2.0$ nA, and  $3.3$ nA for T=300 K, 380 K, and 450 K,
respectively.  At $V_{\rm b} = -5$ V, the corresponding parameters
are: $\alpha=6.3\pm1.3$, $4.6\pm0.5$ and $6.4\pm0.9$,  and $I_{\rm des} =
4.5$  nA, $4.2$ nA and  $6.4$  nA. The data follow the power law very well at all
temperatures and  the  slope is almost
independent of $T$. The temperature effect is largest  at $-7$ V;  for example at 
$I = 2$ nA, $R$ decreases from $700$ s$^{-1}$ to 3 s$^{-1}$   when $T$ is increased from 300 K to 380 K, while at $-5$ V
and $I=8$ nA, $R$ only decreases from $120$ s$^{-1}$ to 40 s$^{-1}$ upon the same temperature change. 

The relationship between $I_{\rm des}$ and $V_{\rm b}$, corresponding to
 a fixed desorption rate
of $R_0 = 4$ s$^{-1}$ or 50\% desorption of H along a
line scan with $s=2$ nm/s is depicted in Fig.~3 for T ranging
from 300 K to 610 K. At a given $T$ and $V_{\rm b}$,  $R >
4$ s$^{-1}$  for values of $I>I_{\rm des}$.   At all $T$,
the data show a minimum in $I_{\rm des}$ at $V_{\rm b} \sim -7$ V, and therefore
 a maximal desorption yield at  this voltage.

Previously we have shown that the room temperature data can be
explained by vibrational heating of the Si-H stretch mode due to 
inelastic  scattering  with the Si-H 5$\sigma$ hole
resonance, and  first principles 
calculations of  $R$ due to this mechanism are in 
quantitative agreement with the experimental
data\cite{StThSaQuHuMuGr98}. In the following we 
use this method to  calculate the $T$ dependence of $R$  and compare it with the experimental data.

 The  first principles method for calculating the 
inelastic current is based on  a high voltage
extension\cite{StQuGr98} of the Tersoff-Hamann model\cite{TeHa85} for the
STM tunnel junction, and has been described in detail
elsewhere\cite{StThSaQuHuMuGr98,StHuThXi98}. We include
 inelastic scattering events with energy transfer
$\pm \hbar\omega_0$, $\pm 2\hbar\omega_0$ and $\pm 3\hbar\omega_0$,  where 
$\hbar \omega_0 = 0.26$ eV is the energy of the Si-H stretch mode.
Energy relaxation due to coupling with silicon phonon modes  is described by 
a current independent energy relaxation rate, $\gamma_T$, which we determine from
experiment\cite{GuLiHi95} in a manner similar to
Ref.~\cite{FoKaLyAv98}.  To obtain  $R$, we
solve the Pauli master equation for the transitions among 
the vibrational  levels of the H potential well, and assume that
desorption occurs when the energy of the H atom exceeds the desorption
energy $E_{\rm des}=3.36$~eV, corresponding to a truncated
harmonic potential well  with 13 levels.

In the calculation of the inelastic current a correction is made for the
difference in the  Fermi level, $\varepsilon_F^s$ and  the band bending,
$\Phi$, between  the slab and the sample. This is carried out by using an offset between the  sample bias of the slab model
$ \tilde{V}_{\rm b}$ and the experimental $V_{\rm b}$\cite{StQuGr98}
\begin{equation}
   eV_{\rm b}= e\tilde{V}_{\rm b} +
\tilde{\varepsilon}_F^s-\tilde{\Phi}- (\varepsilon_F^s-\Phi),
\label{eq:vchang}
\end{equation}
where the values with a tilde are slab quantities. Both  $\varepsilon_F^s$
and $\Phi$  depend on the sample temperature.  However,  since the
changes in $\Phi$ and $\varepsilon_F^s$ nearly cancel each other, the
offset, $ V_{\rm b}-\tilde{V}_{\rm b}=0.30\pm0.05$ V, is almost constant
for the temperatures considered in this work.  
The dependence of the inelastic current on $T$ due to changes
in $\Phi$ and $\varepsilon_F^s$ can therefore be neglected.  

We first consider the effect of a $T$ dependence of $\gamma_T$.  The
relaxation of the 
Si-H stretch mode due to phonon-phonon interaction is expected to
involve three quanta of the Si-H 
bending mode, $\hbar \omega_{\rm b}=0.078$ eV,  and a Si phonon
$\hbar \omega_{\rm Si}=0.026$ eV \cite{GuDuChHi90,FoKaLyAv98}. The
$T$ dependence of this interaction is described by\cite{PeAv97}
\begin{equation}
\gamma_T = \gamma_0 \{[1+ n_T(\omega_{\rm
 b})]^3[1+ n_T(\omega_{\rm Si})]- n_T(\omega_{\rm b})^3
 n_T(\omega_{\rm Si})\},
\label{eq:gamt}
\end{equation}
where $ n_T(\omega) = 1/(\exp(\hbar \omega/kT)-1) $ is the
 Bose-Einstein occupation number.  Using  the
 experimental room temperature relaxation rate, $\gamma_{300} = 10^8$
 s$^{-1}$ and Eq.~(\ref{eq:gamt}) the relaxation rate at $T=0$ K is
 determined to be, $\gamma_0=5 \times 10^7$ s$^{-1}$. At 600 K the relaxation rate obtained from Eq.~(\ref{eq:gamt})
 is  , $\gamma_{600}= 3 \times 10^8$ s$^{-1}$, which is  3 times
 larger  than the value at room temperature.  

We note that lateral diffusion of the excitation into
 the H overlayer through incoherent exciton motion is not important
 for the present system. This is due to the fact that  inelastic scattering with an
 energy transfer of  $2 \hbar\omega_{0}$ dominates the desorption
 process and the  lateral diffusion rate for excitations with $n>1$ is
 very low due to the anharmonicity of
 the Si-H bond potential\cite{PeAv97,St98}.

The dashed curves in Fig.~2 show the result of this model for the
dependence of $R$ on $I$  for $V_{\rm b} = -7$ V and $V_{\rm b}
= -5$ V and $T=300$ K, 380 K and 450 K. We note that there are no
adjustable parameters in the model. For the present system  inelastic scattering events  with
energy transfer $2\hbar \omega_0$ give the dominant contribution to $R$  in the  temperature and voltage range investigated.  Since there
are 13 levels in the
truncated harmonic potential well of  the H
atom\cite{StThSaQuHuMuGr98} it can be shown that  $R\sim (I_2/\gamma_T)^{13/2}$\cite{SaPePa94} where $I_2$ is the inelastic current with energy transfer $2\hbar \omega_0$.  From Eq.~(\ref{eq:gamt}) we  have $\gamma_T= 1$, $1.3$, and $1.7 \times 10^8 {\rm s}^{-1}$ at $T=300$, $380$, and  $450$ K, respectively. From this model we obtain $R(300)/R(380) \sim 1.3^{6.5} \sim 5$ and $R(300)/R(450) \sim 30$.
At  $V_{\rm b}=-5$ V this is in good agreement with 
 the measured $T$  dependence,  whilst the  measured $T$  dependence
is almost two  orders of magnitude larger  at
$V_{\rm b}=-7$ V. The dashed curves in Fig.~3  show the calculated  bias dependence of  $I_{\rm des}$
for temperatures, 300 K, 380 K, 450 K, 570 K, and 610K. 
The minimum in $I_{\rm des}$ occurring  at $-7$ V is reproduced by the
model and is related to the position of the energy level of the
5$\sigma$ hole resonance\cite{StThSaQuHuMuGr98}. For all voltages $R$  decreases with $T$. For   $V_{\rm b}>   - 5$ V there is quantitative agreement between the
theoretical model and the experimental data.  However, for $ V_{\rm b} < -5$ V there is an additional
temperature dependence not accounted for  by the $T$ dependence of $\gamma_T$.

As already discussed for the present system there are no significant temperature
effects due to band bending and   lateral
diffusion  of the vibrational excitation, and  calculations of the tunnel current
 using the appropriate Fermi-Dirac occupation factors show that the thermal
current is negligible. At the elevated temperatures we still have atomic
resolution for imaging and lithography(see Fig.~1), implying that  the
temperature does not affect the tip shape.  At elevated
temperatures the  deposition of H atoms from the tip to
the surface becomes important\cite{SaThNaAo97}, but this effect is not
large enough to  change $R$ by a factor of 100.

Instead we consider the effect of a $T$ dependent lifetime of the $5\sigma$ hole
resonance. The time evolution of the wave function,
$\phi_{5\sigma}$, is given by
\begin{equation}
\phi_{5\sigma}(t)= \sum_{\mu} \langle \mu |5\sigma \rangle \psi_{\mu}(0) e^{-i \varepsilon_{\mu} t/\hbar} e^{-t/2\tau_{\rm ph}},
\end{equation}
where $\psi_{\mu}$ are  eigenstates of the adsorbate+substrate system
with eigenvalue, $\varepsilon_{\mu}$, and due to electron-phonon
couplings they have a  finite
lifetime, $\tau_{\rm ph}$, which we take to be independent of
$\mu$. If  the $5\sigma$ resonance is approximated by a Lorentzian
line shape we obtain
\begin{equation}
\phi_{5\sigma}(t)= \phi_{5\sigma}(0)  e^{-i \varepsilon_{5\sigma}
t/\hbar} e^{-(1/2\tau_{\rm e}+1/2\tau_{\rm ph}) t},
\end{equation}
where, $\varepsilon_{5\sigma}$, is the resonance energy and 
 $\tau_{\rm e}$ is the lifetime  due to electronic coupling with  substrate
 eigenstates. The lifetime of the resonance may now change with
 temperature due a $T$ dependence of both $\tau_{\rm ph}$ and
 $\tau_{\rm e}$. The $T$ dependence of $\tau_{\rm ph}$ for different
 eigenstates of Si has been measured in
 Ref.~\cite{LaAlCa86,LaGaViCa87}, and  $\hbar/2\tau_{\rm ph} \sim 0.1$
 eV at room temperature and $\hbar/2\tau_{\rm ph} \sim 0.2$ eV at 600 K\cite{LaAlCa86,LaGaViCa87}. We also expect a
$T$ dependence of $\tau_{\rm e}$, since the electronic coupling between H and the Si surface depends on the position of the Si atoms. 

The energy dependent probability of inelastic scattering,
 $P_n(\varepsilon)$, with transfer
of energy, $n \hbar \omega$,  is  for a Lorentzian resonance given by\cite{SaPePa94}
\begin{equation}
P_n(\varepsilon,\Delta_T) \propto \frac{\Delta_T}{[(\varepsilon-\varepsilon_{5\sigma})^2
+ \Delta_T^2]^{n+1}},
\label{eq:pn}
\end{equation}
where $\Delta_T=1/2\tau_{\rm ph}+1/2\tau_{\rm e}$ is the Half Width at
Half Maximum(HWHM) of the resonance at temperature $T$. From $P_n$
the $T$ dependent inelastic currents can be calculated
using
\begin{equation}
I_n(\varepsilon,T) \approx I_n(\varepsilon,0)
\frac{P_n(\varepsilon,\Delta_T)}{P_n(\varepsilon,\Delta_0)},
\label{eq:IT}
\end{equation}
where  $I_n(\varepsilon,0)$ is the first principles value for the
$T=0$ K contribution to the inelastic current of tunneling holes with
energy  $\varepsilon$, and 
$\Delta_0=0.6$ eV, $\varepsilon_{5\sigma}-\varepsilon_F^s \sim -6$ eV(including band bending) are the 
parameters at $T=0$\cite{StThSaQuHuMuGr98}. 

   We determine the $T$ dependent width, $\Delta_T$, 
by fitting $R$ calculated using $I_n(\varepsilon,T)$ and $\gamma_T$ to
the measured  values at $V_{\rm b} = -7$ V.  The
fitted values of $\Delta_T$ are plotted as function of $T$  in Fig.~4, and  solid curves in Fig.~2
show  the  corresponding $T$ dependent  values of $R$ as function of $I$ for $V_{\rm b}
= -5$ V and $V_{\rm b} = -7$ V.  At $V_{\rm b}= -5$ V the dashed curves
and the solid curves coincide, and the $T$ dependence of $R$ 
  mainly stems from $\gamma_T$, whilst  there is a
strong additional $T$ dependence of $R$ from $\Delta_T$ at  $V_{\rm
b}= -7$ V. This behaviour can be understood using  $R\sim
(I_2/\gamma_T)^{6.5}$ and taking into account the $T$ dependence of
$I_2$ from Eq.~(\ref{eq:IT}). Using Eq.~(\ref{eq:pn}) it is observed
that $I_2
\propto \Delta_T$ for $|V_{\rm b}| \ll
|\varepsilon_F^s-\varepsilon_{5\sigma}|$ and $I_2  \propto
\Delta_T^{-5}$  for $|V_{\rm b}| \ge 
|\varepsilon_F^s-\varepsilon_{5\sigma}|$. At $V_{\rm b}= -7$ V this gives a 
$T$ dependence of $R(300)/R(380) \sim (1.3 \times 1.13^5)^{6.5} \sim 10^2$ and
$R(300)/R(450) \sim 10^4$  in good agreement with the measured
data. At $V_{\rm b}= -5$ there is a weak $T$ dependence of $R$ upon
$\Delta_T$ since $V_{\rm b}| < |\varepsilon_F^s-\varepsilon_{5\sigma}|$
causes  $I_2$ to be  nearly independent of $T$ and
 the $T$ dependence of $R$ therefore mainly stems from $\gamma_T$. 
The solid curves  plotted in Fig.~3 as obtained from the present model
are in good agreement with the experimental data for the whole bias
range from $-10$ V to $-4$ V.   It is observed that the curves become
 more flat with increasing temperatures in good agreement with experimental data.
 At positive bias inelastic scattering with the Si-H 6$\sigma^*$ is   the dominant desorption mechanism for $|V_{\rm b}| \le 3$. In this bias regime
$|V_{\rm b}|  \ll |\varepsilon_F^s-\varepsilon_{6\sigma^*}| \sim 6$ eV
and at this condition  the main   $T$ dependence of $R$ must come from $\gamma_T$ in good agreement with Ref.~\cite{FoKaLyAv98}.  

In order to investigate the origin of the 
$T$-dependence of $\Delta_T$ we model $\Delta_T$ by\cite{LaAlCa86}
\begin{equation}
\Delta_T = \Delta_0 + a (e^{\Theta/kT}-1)^{-1}.
\label{eq:fit}
\end{equation}
From a least-squares fit to the data in Fig.~4 we obtain
$\Delta_0=0.49$ eV, $a=0.36$ eV, and  $\Theta=402$ K. The $T$
dependence 
of $\Delta_T$ is quite strong and almost linear for $T> 300$ K with the slope
$a/\Theta \sim 1$ meV/K. The measured dependence upon  $T$ of
$\Delta_{\rm ph}$ for silicon is  in the range $ 0.1$-- $0.4$
meV/K\cite{LaAlCa86,LaGaViCa87}, and a  major part of the $T$
dependence must therefore come  from
$\Delta_{\rm e}$. This dependence must be related to a change in the
electronic coupling between the adsorbate and the surface upon
excitations of phonons with frequency $\hbar \Omega \sim \Theta$. However,
the fit in Eq.~(4) is quite insensitive to the value of $\Theta$ and
from the present data 
we can only state that $\hbar\Omega < 0.1$ eV. It is  therefore not possible to distinguish whether the   $T$ dependence of $\Delta$ is from Si
phonons or Si-H transverse vibrations.

In summary we have presented experimental data of the 
desorption rate of STM induced
desorption of H from the Si(100)-H(2$\times$1) surface as a function
of temperature, and found a decreasing  desorption rate as function of
the  temperature. The desorption mechanism is related to 
vibrational heating of the H atom by holes  inelastic scattering  with
the Si-H 5$\sigma$  hole
resonance, and the temperature dependence is explained by  both a decreasing 
vibrational life time of the Si-H stretch mode and a decreasing
electronic life time  of the Si-H $5\sigma$
hole resonance with increasing temperature.

We would like to acknowledge B. Y. Hu, F. Grey, A. P. Jauho, and U. Quaade for many
valuable discussions. This work was supported by  the Japanese Science and
Technology Agency and the Danish research councils(STVF) through 
talent project No. 9800466. The use of Danish national computer
resources was supported by the Danish Research Councils.

\begin{figure}
\begin{center}
\leavevmode
\epsfxsize=85mm
\epsffile{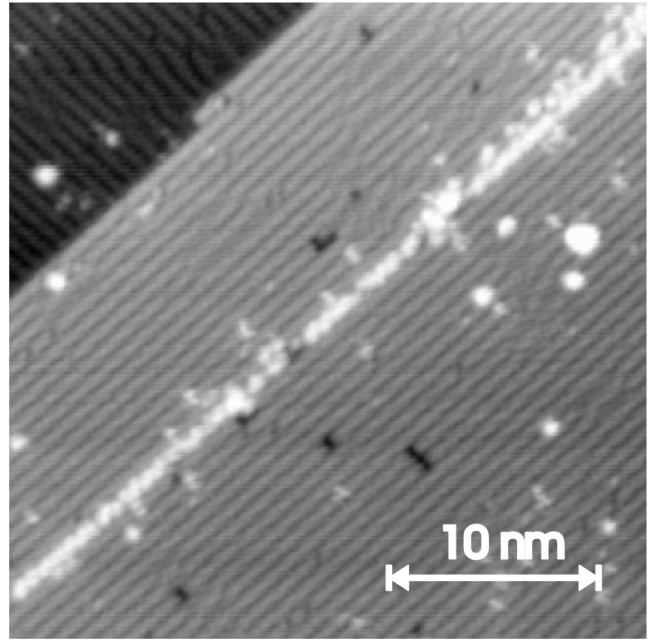}
\end{center}
\caption{ Scanning tunneling micrograph of exposed silicon
dangling bonds as a result of hydrogen (H) desorption on a H
terminated Si(100)-(2$\times$1) reconstructed surface of the $n$-type
sample at a substrate temperature of 530 K and a line scan
with a speed of 2 nm/s, a sample bias ($V_{\rm b}$) of $-7$ V and a
tunnel current ($I$) of 8.0 nA. The filled-state image was
recorded at $V_{\rm b}=-1.7$ V and $I=0.2$ nA.} 
\end{figure}

\begin{figure}
\begin{center}
\leavevmode
\epsfxsize=85mm
\epsffile{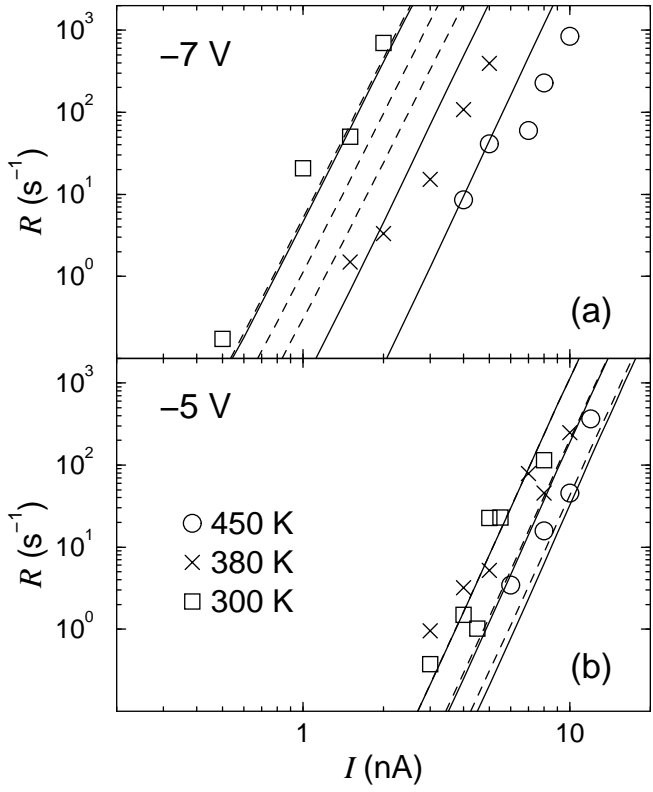}
\end{center}
\caption{STM induced desorption rate ($R$) 
as function of tunnel current ($I$) for temperatures
300 K, 380 K and 450 K for  sample biases  of
(a) $-7$ V and (b) $-5$ V.  Curves show theoretical calculations of
$R$ using only a temperature dependent vibrational lifetime(dashed)
and combined with a temperature dependence of the electronic lifetime(solid).} 
\end{figure}

\begin{figure}
\begin{center}
\leavevmode
\epsfxsize=85mm
\epsffile{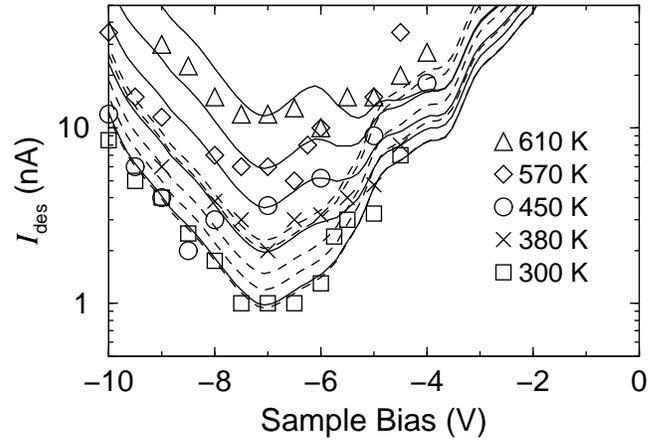}
\end{center}
\caption{ The tunnel current ($I_{\rm des}$) as function of sample bias 
($V_{\rm b}$) which gives rise to 50\% desorption of H from a 
Si(100)-H(2$\times$1) surface along a line scan of 2 nm/s by an STM tip.
Results are plotted for temperatures between 300 K, 380 K, 450 K, 570
K, and  610 K. Curves show theoretical calculations(see caption of Fig.~2).}
\end{figure}

\begin{figure}
\begin{center}
\leavevmode
\epsfxsize=85mm
\epsffile{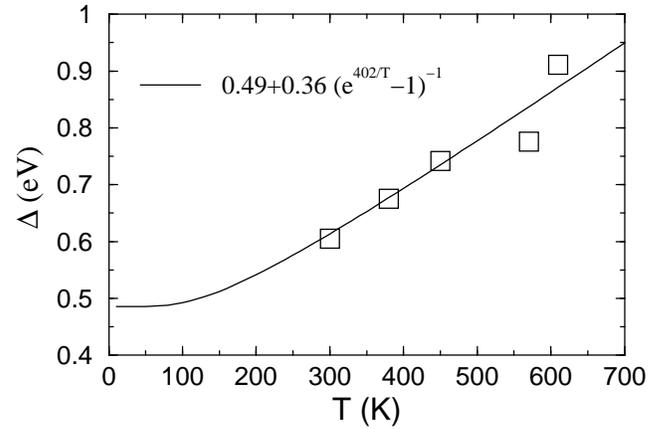}
\end{center}
\caption{ Squares show temperature dependent  values of $\Delta$
obtained by fitting theoretical desorption rates to the experimental
values. The solid curve shows a least-squares fit of Eq.~(\ref{eq:fit}) to the
data points.}
\end{figure}

\end{multicols}  
\end{document}